\documentclass[a4paper,12pt]{article}
\usepackage{graphicx,amssymb,color}
\usepackage{epsfig}
\usepackage{color}

\begin{document}

\centerline{\LARGE \bf Doubly nonlocal reaction-diffusion equations}

\medskip

\centerline{\LARGE \bf and the emergence of species}

\vspace*{1cm}

\centerline{\bf M. Banerjee$^1$, V. Vougalter$^2$, V. Volpert$^3$}

\vspace*{0.5cm}

\centerline{$^1$ Department of Mathematics and Statistics,
Indian Institute of Technology Kanpur}
\centerline{Kanpur 208016, India}

\centerline{$^2$ Department of Mathematics, University
of Toronto, Toronto, Ontario, M5S 2E4, Canada}

\centerline{$^3$ Institut Camille Jordan, UMR 5208 CNRS, University Lyon 1}
\centerline{69622 Villeurbanne, France}

\vspace*{1.5cm}

\noindent
{\bf Abstract.}
The paper is devoted to a reaction-diffusion equation with doubly nonlocal nonlinearity arising
in various applications in population dynamics. One of the integral terms corresponds to the nonlocal
consumption of resources while another one describes reproduction with different phenotypes.
Linear stability analysis of the homogeneous in
space stationary solution is carried out. Existence of travelling waves is proved in the case
of narrow kernels of the integrals. Periodic travelling waves are observed in numerical simulations.
Existence of stationary solutions in the form of pulses is shown, and transition from periodic
waves to pulses is studied. In the applications to the speciation theory, the results of this work signify that
new species can emerge only if they do not have common offsprings.
Thus, it is shown how Darwin's definition of species as groups of morphologically similar individuals
is related to Mayr's definition as groups of individuals that can breed only among themselves.\\


\vspace*{1cm}


\section{Nonlocal equations in population dynamics}


Nonlocal reaction-diffusion equations arise in various applications. In population dynamics they are widely used in order to describe
nonlocal consumption of resources \cite{Britton, GVA1, Gourley2, v2014} or breeding with different phenotypes \cite{AV3, DV, DMV}.
%
%
The roles of these nonlocal terms are quite different, from the biological and from the modelling points of view. In this work we will consider both
of them at the same time and will study their mutual influence.
Their combination will allow us to make some important conclusions about the emergence of biological species.

We consider the nonlocal reaction-diffusion equation

\begin{equation}\label{q4}
   \frac{\partial u}{\partial t} = D \frac{\partial^2 u}{\partial x^2} + a (S(u))^2 (1 - J(u)) - b u ,
\end{equation}
where $D$, $a$ and $b$ are some positive constants,

$$ S(u) = \frac{1}{2h_1} \int_{-\infty}^{\infty} \psi(x-y)u(y,t)dy , \;\;\;\;
\psi(z) = \left\{\begin{array}{cc}
1, & |z|\leq h_1\\
0, & |z|> h_1
\end{array}  \right. , $$

$$ J(u) = r(h_2) \int_{-\infty}^{\infty} \phi(x-y)u(y,t)dy ,   \,\,\,\,\,
\phi(z) = \left\{\begin{array}{cc}
1, & |z|\leq h_2\\
0, & |z|> h_2 \end{array}\right. . $$
We take the kernels of the integrals in the form of step-wise constant functions in order
to simplify analysis and simulations. Other kernels can also be considered.
We will set  $r(h_2)= 1 /(2h_2)$ or $r(h_2)= 1$. In the first case,
in the limit of small $h_2$ we get $J(u)=u$ similar to the local consumption of resources
in the logistic equation. In the case of asymptotically large $h_2$ we put
$r(h_2)=1$. Then in the limit of large $h_2$ we obtain global consumption of resources with the
integral $I(u) = \int_{-\infty}^{\infty}u(y,t)dy$ in the consumption term.

Various particular cases of this equation are studied in the literature (see \cite{AADV, v2014, VP} and the references therein).
In the limit of small $h_1$ and $h_2$ we obtain the classical reaction-diffusion equation

\begin{equation}\label{in2}
   \frac{\partial u}{\partial t} = D \frac{\partial^2 u}{\partial x^2} + a u^2 (1 - u) - b u .
\end{equation}
If $b < a/4$, then the nonlinearity has three zeros, $u_+ = 0$, $u_0 = (1 - \sqrt{ 1 - 4{b}/{a}})/2$
and $u_- =  (1 + \sqrt{ 1 - 4{b}/{a}})/2$. In this case there is a travelling wave solution of this equation,
$u(x,t) = w(x-ct)$ with the limits $w(\pm \infty) = u_\pm$. It exists for a unique value of $c$ and it is
globally asymptotically stable (see \cite{v2014, VVV} and the references therein). This equation also has stationary solutions in the form
of pulses, that is positive solutions with zero limits at infinity. Such solutions are unstable.

If $h_1>0$ and $h_2=0$, then equation (\ref{q4}) corresponds to the local consumption of resources and breeding with possibly
different phenotypes (see Appendix):

\begin{equation}\label{in3}
   \frac{\partial u}{\partial t} = D \frac{\partial^2 u}{\partial x^2} + a (S(u))^2 (1 - u) - b u .
\end{equation}
Similar to the previous case, existence and stability of travelling waves for this equation is proved \cite{AV3, DV, DMV}.
The next particular case of equation (\ref{q4}) is $h_1=0$ and $h_2>0$:

\begin{equation}\label{in4}
   \frac{\partial u}{\partial t} = D \frac{\partial^2 u}{\partial x^2} + a u^2 (1 - J(u)) - b u .
\end{equation}
Contrary to the previous equation, the maximum principle is not applicable here, and the wave existence
is proved only for $h_2$ sufficiently small \cite{ADV1, ADV2} or
in some other special cases \cite{ACR}. Propagation of periodic waves is observed in numerical simulations for the values of $h_2$ greater than some
critical value $h_2^*$ for which the homogeneous in space solution $u=u_-$ loses its stability resulting
in appearance of stationary periodic solutions.

In the monostable stable case ($u^2$ in the production term
is replaced by $u$), existence of waves is proved for all speeds greater than or equal to the minimal speed
\cite{AC, ABVV, BNPR, Gourley}, stability of waves is studied in \cite{VV1}, their dynamics in \cite{ADV1, Aydogmus, GVA1, nadin, PG}.
Systems of equations are studied in \cite{Bayliss, Segal, v2014, Zwolenski}.

Equation (\ref{in4}) has stationary solutions in the form of pulses for all $h_2$ sufficiently
large ($r(h_2)=1$) \cite{VV2}. Numerical simulations show that these pulses are stable \cite{BRV, v1, v2},
though their stability is not proved analytically.

Thus, equation (\ref{in4}) has travelling waves propagating with a constant speed and profile for $h_2$
sufficiently small. Periodic waves are observed for intermediate values of $h_2$ and pulses for $h_2$
sufficiently large. Transition from periodic waves to pulses occurs through a global bifurcation where
the speed of the periodic wave converges to zero and the peaks of the wave form the pulses \cite{v2}.

Equation (\ref{in4}) is a limiting case of equation (\ref{q4}) as $h_1$ converges to $0$. In this work we will study
how dynamics of solution is influenced by the integral $S(u)$ for positive $h_1$. The linear stability analysis
of the homogeneous in space solutions is presented in Section 2. Dynamics of pulses and waves is discussed in Section 3.
The derivation of the model and positiveness of solutions are discussed in the appendix.

One of the important applications of nonlocal reaction-diffusion equations concerns the theory of speciation.
Nonlocal consumption of resources allows the description of the emergence of biological species \cite{GBV, GVA1, GVA2, VP}.
The integral $S(u)$ in the reproduction term is determined by the relation between the phenotypes of parents.
We will show that speciation can occur only if this distribution is sufficiently narrow. This is related
to Mayr's definition of species as a group of individuals that can breed only among themselves.
We discuss these questions in Section 4.



\setcounter{equation}{0}

\section{Linear stability analysis}

\subsection{Nonlocal consumption}

%
%

We begin with the stability analysis of a homogeneous in space stationary solution.
Linearizing equation (\ref{q4}) about $u=u_* (=u_\pm, u_0)$, we obtain the eigenvalue problem:


\begin{eqnarray}\label{evproblem}
D v^{''}+2 a u_*(1-u_*) S(v) 
- a u_*^2 J(v) 
-bv = \lambda v .
\end{eqnarray}

\begin{figure}[htbp]
\centerline{\includegraphics[scale=0.3]{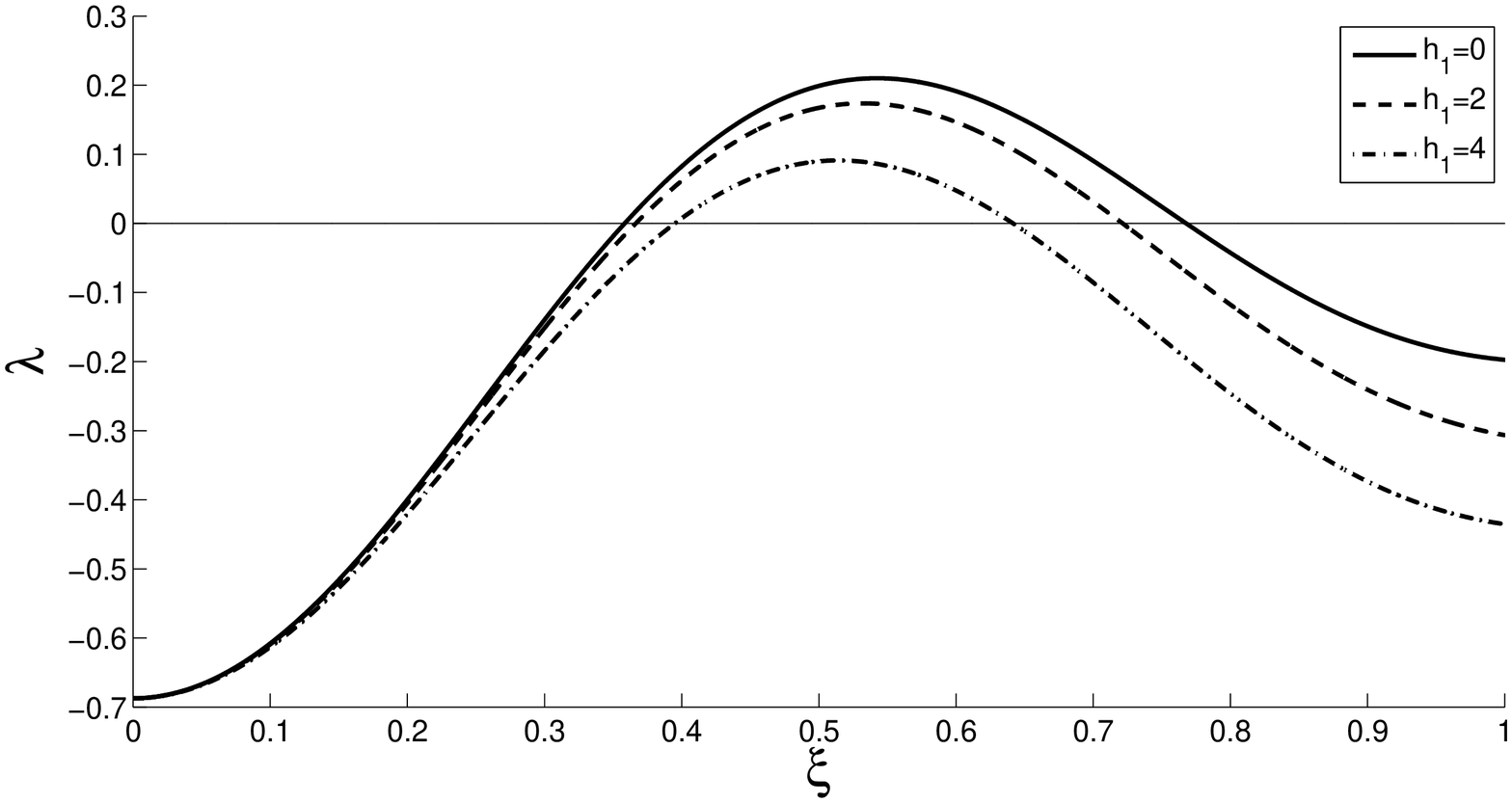}
\includegraphics[scale=0.3]{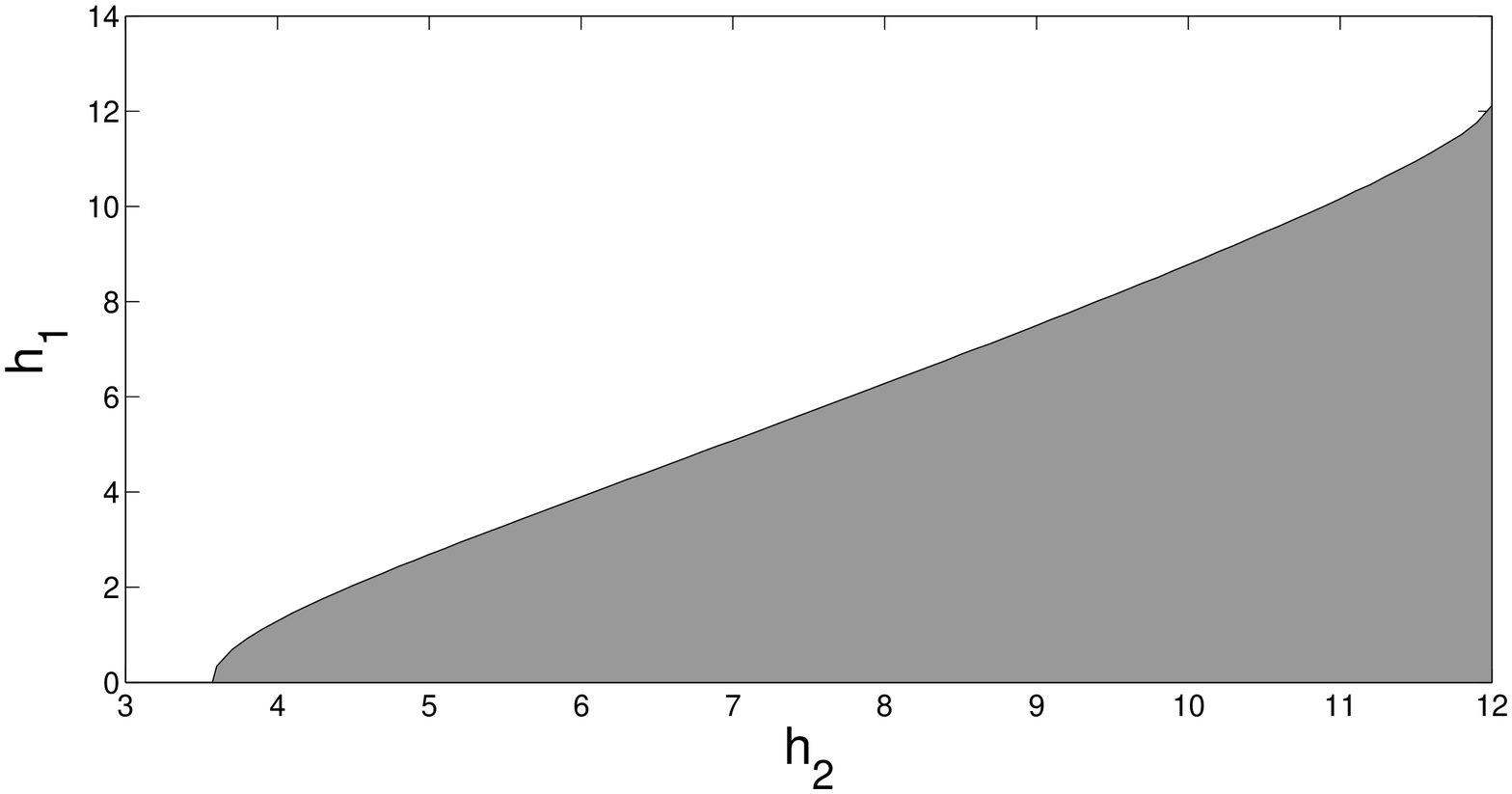}}
\caption{Linear stability analysis. The eigenvalue $\lambda$ in
(\ref{ev}) for $D=0.2$, $a=1$, $b=0.1$, $h_2=5$ and for different
values of $h_1$ (left). Stability region (white) and instability
region (grey) on the $(h_2,h_1)$ plane.}
\label{diagram}
\end{figure}

Taking the Fourier transform, we find the spectrum:

\begin{equation}
\label{ev}
\lambda = - D \xi^2 + 2 a u_*(1-u_*)\tilde \psi (\xi) -  a u_*^2\tilde \phi(\xi) - b ,
\end{equation}
where
\begin{eqnarray*}
\tilde \psi (\xi)\,=\,\frac{\sin(\xi h_1)}{\xi h_1} \; ,\,\,\,
\tilde \phi (\xi)\,=\,\frac{\sin(\xi h_2)}{\xi h_2} \; .
\end{eqnarray*}
In the limit of small $h_1$ and $h_2$, which corresponds to the local reaction-diffusion equation,
we obtain that the solution $u=u_-$ is always stable. Since $\tilde \phi (\xi) \leq 1$,
then the function $\tilde \phi$ moves the spectrum to the right, and it can destabilize the solution.
Hence nonlocal consumption of resources can lead to the instability of the homogeneous in space
stationary solution. This result is known from the numerous previous works \cite{Britton, GVA1, GVA2, Gourley2}.
The function $\tilde \psi \leq 1$ moves the spectrum to the left, and it stabilizes the solution.

The eigenvalue $\lambda$ as a function of $\xi$ is shown in Figure \ref{diagram} (left). It is negative for $\xi=0$
and positive on some interval of $\xi$ providing the instability of the stationary solution. The interval
where it is positive decreases with the increase of $h_1$. Stability boundary is given by the values of parameters
for which the maximal value of $\lambda(\xi)$ equals $0$.
Stability and instability regions on the plane $(h_2, h_1)$
are presented in Figure \ref{diagram} (right). The instability region is below the curve. Increase of $h_2$
destabilizes the solution while increase of $h_1$ stabilizes it.



%



\subsection{Global consumption}

The integral $I(u)= \int_{-\infty}^\infty u(x,t)dx$ is well defined only for functions $u(x,t)$ integrable on the whole axis.
Therefore in order to study the emergence of pulses in the case of global consumption, we will consider a similar equation

\begin{equation}\label{h1}
   \frac{\partial u}{\partial t} = D \frac{\partial^2 u}{\partial x^2} + a u^2 (1 - I_0(u)) - b u ,
   \;\;\;\; I_0(u) = \int_0^L u(y,t)dy
\end{equation}
on a bounded interval $0 < x < L$ with the no-flux boundary conditions: $x=0,L : \partial u/\partial x = 0$.
To simplify the presentation, we consider the particular case where $S(u)=u$ ($h_1=0$).
%

%
%

We look for homogeneous in space stationary solutions of equation (\ref{h1}).
If $b < a/(4L)$, then this problem has three constant solutions, $w=0$,
and two solutions of the equation

\begin{equation}\label{3}
 a w (1 - I_0(w)) = b .
\end{equation}
 We denote them by $w_1$ and $w_2$ assuming that $w_1 < w_2$.

Consider next the eigenvalue problem for the equation linearized about a constant solution $u_*$:

\begin{equation}\label{4}
  D u'' + 2 a u_* (1- I_0(u_*)) u - b u - a u_*^2 I_0(u) = \lambda u , \;\;\;
  u'(0) = u'(L) = 0 .
\end{equation}
Taking into account (\ref{3}) we can write it as follows:

\begin{equation}\label{5}
  D u'' +  b u - a u_*^2 I_0(u) = \lambda u , \;\;\;
  u'(0) = u'(L) = 0 .
\end{equation}
We will search its solutions in the form

$$ u(x) = \cos(n \pi x/L) , \;\;\; n=0,1,2,... $$
Then we get

$$ \lambda_0 = b - a u_*^2 L , \;\;\;
\lambda_n = - D (n \pi/L)^2 + b , \;\; n=1,2,... $$
Hence the presence of the integral term influences only the eigenvalue $\lambda_0$.
From (\ref{3}) we get

$$ \lambda_0 = a u_* (1 - 2 L u_*) . $$
If equation (\ref{3}) has two solutions, then $\lambda_0 > 0$ for $u_*=w_1$ and
$\lambda_0 < 0$ for $u_*=w_2$.

Thus, the problem linearized about solution $w_2$ has negative eigenvalue $\lambda_0$.
The eigenvalue $\lambda_1$ can be negative or positive. If it is negative, this solution is stable,
otherwise it is unstable and another solution bifurcates from it.
We can consider $D$ as bifurcation parameter with the critical value $D^* = b L^2/\pi^2$.
If $D < D^*$, then a non-homogeneous stable solution emerges.
Since the eigenfunction $\cos(\pi x/L)$ corresponding to the eigenvalue $\lambda_1$ has its
extrema at the boundary, then the emerging solution has also its maximum at the boundary of the interval.
If we consider a double interval, then this solution corresponds to the pulse solution.





\setcounter{equation}{0}

\section{Waves and pulses}

\subsection{Wave propagation}

Travelling wave solution of equation (\ref{q4}) is a solution $u(x,t)=w(x-ct)$, where the function $w(\xi)$
satisfies the equation

\begin{equation}\label{wp1}
  D w'' + c w' + a (S(w))^2 (1 - J(w)) - b w = 0 , \;\;\;\; w(\pm \infty) = u_\pm .
\end{equation}
The constant $c$ is the wave speed. It is a priori unknown and should be found as a solution of the problem.
We suppose here that $r(h_2) = 1/(2h_2)$. In the limit of small $h_2$ we have $J(w)=w$, and we obtain equation
(\ref{in3}) for which the existence of waves is proved \cite{AV3, DV}. Therefore we can expect that the waves also exist for
all $h_2$ sufficiently small.

\medskip

\noindent
{\bf Theorem 3.1.}
{\em For any $h_1>0$ and for all $h_2>0$ sufficiently small problem (\ref{wp1}) has a monotonically decreasing
solution for a unique value of $c$.}

\begin{figure}[htbp]
\centerline{\includegraphics[scale=0.5]{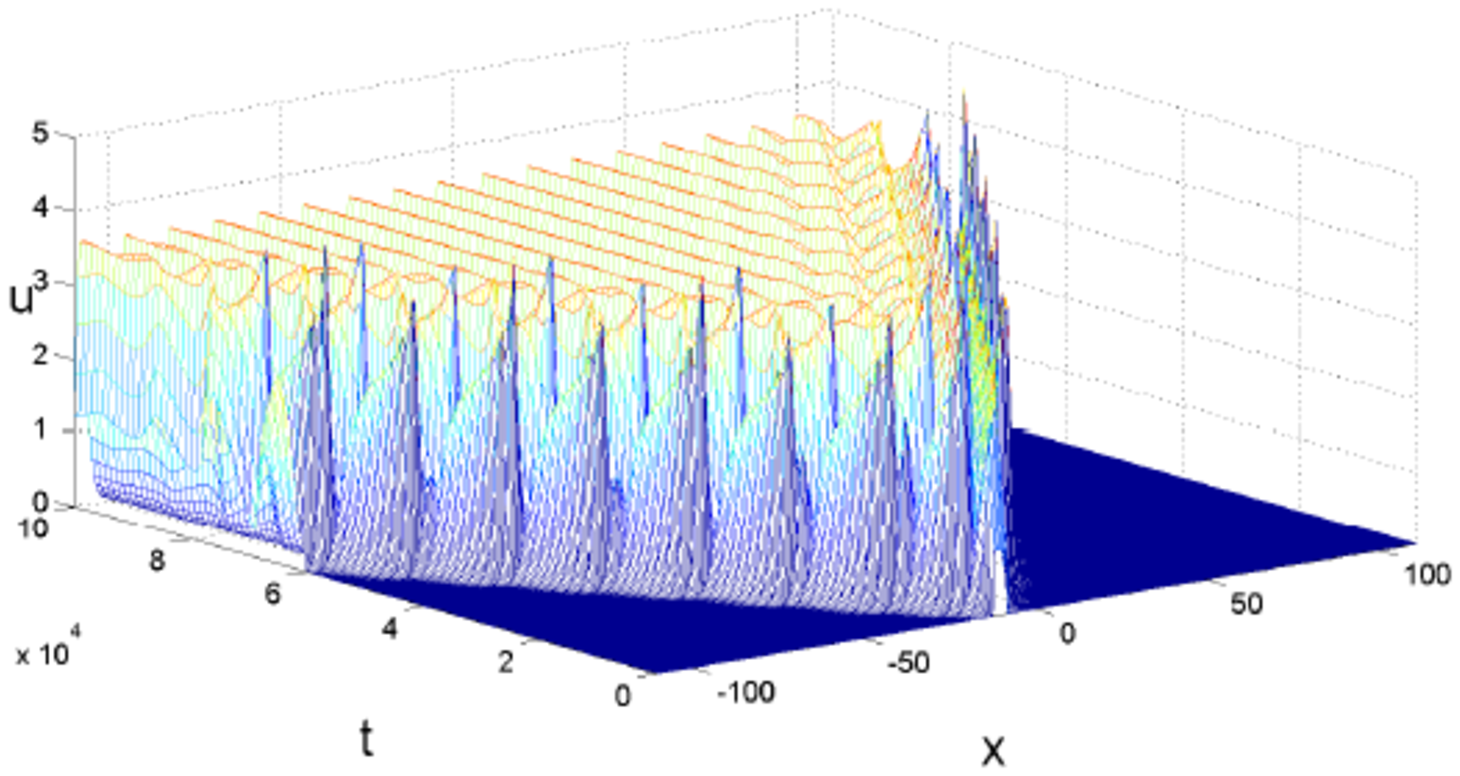} \hspace*{-1cm}
 \includegraphics[scale=0.32]{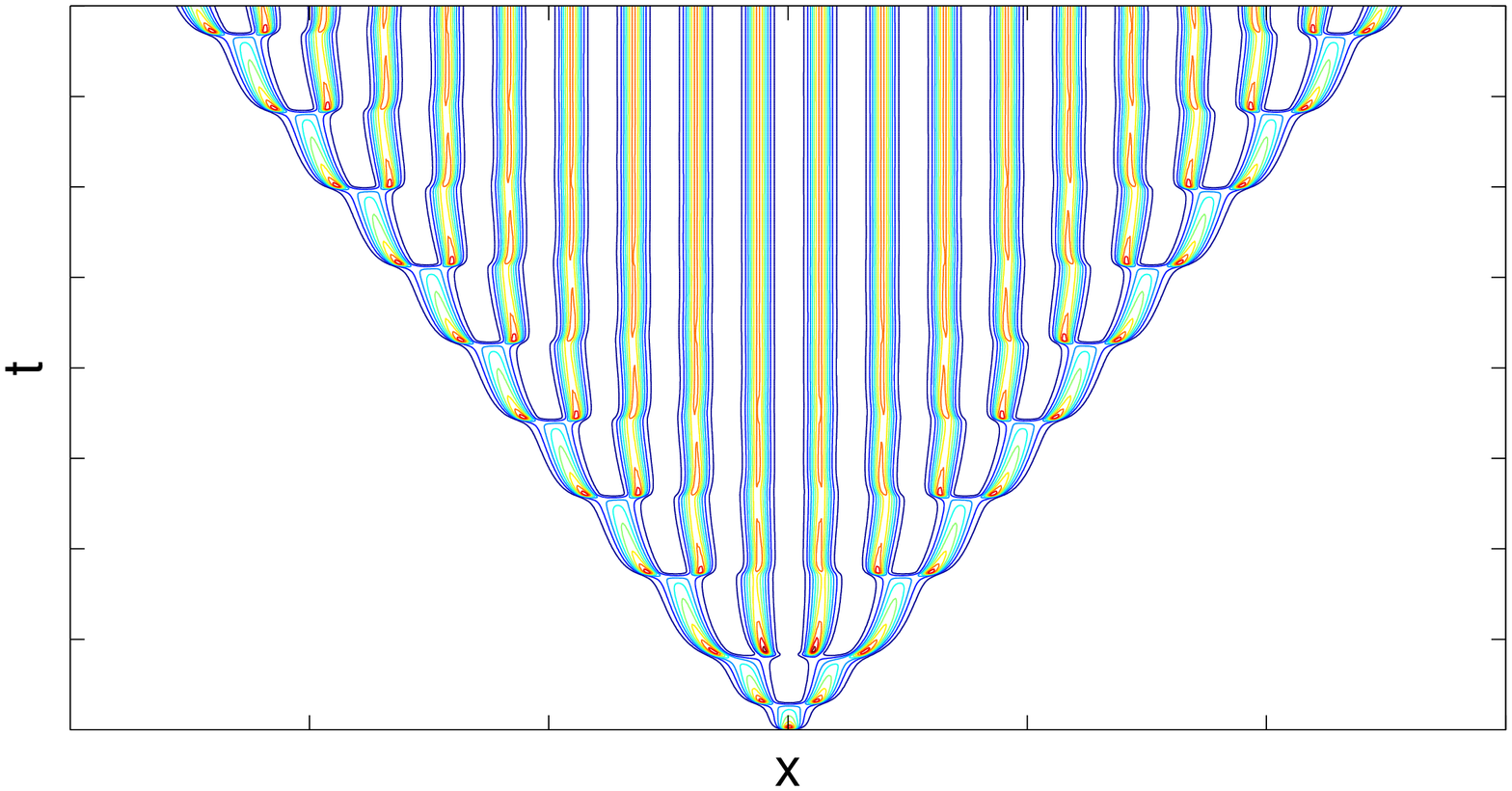}}
 \caption{Propagation of periodic waves for equation (\ref{q4}). Solution $u(x,t)$ as a function of
 two variables (left). The same solution as levels lines of the function $u(x,t)$ on the $(x,t)$ plane.}
 \label{wave}
\end{figure}

\medskip

\noindent
The proof of this theorem is similar to the proof of the corresponding theorem in \cite{ADV1, ADV2}. We use here the existence of solutions for
the limiting equation (\ref{in3}) and the implicit function theorem. The spectrum
of the operator linearized about the solution is completely in the left-half plane except for a simple zero eigenvalue \cite{DMV}.
Therefore the operator is invertible on the subspace orthogonal to the corresponding eigenfunction. This perturbation method
is conventionally used for travelling waves \cite{v2014}.

The wave existence is proved only for sufficiently small values of $h_2$. For larger $h_2$ we can study wave propagation
in numerical simulations. When we increase $h_2$, the wave becomes non-monotone with respect to $x$. For $h_2$ sufficiently large,
the homogeneous in space solution $u=u_-$ becomes unstable resulting in appearance of a stable stationary space periodic solution.
For such values of $h_2$ we observe propagation of a periodic wave where its speed and its profile change periodically in time
(Figure \ref{wave}). This behavior is qualitatively similar to the behavior of solutions of equation (\ref{in4}).


\subsection{Existence of pulses}

Equation

\begin{equation}\label{non1}
  \frac{\partial u}{\partial t} = D \frac{\partial^2 u}{\partial x^2} + a u^2 (1-I(u)) - b u ,
\end{equation}
where

$$ I(u) = \int_{-\infty}^\infty  u(x,t) dx  $$
is the limiting case of equation (\ref{in4}) as $h_2 \to \infty$ ($r(h_2)=1$).
It can be easily verified that it has a pulse solution if $a > a_c$, where the critical value $a_c$ depends on $D$ and $b$ \cite{v2014}.
Equation (\ref{non1}) is formally obtained from equation (\ref{q4}) in the limit of small $h_1$
and large $h_2$.  Existence of pulses for equation (\ref{q4}) is given by the following theorem.

\medskip

\noindent
{\bf Theorem 3.2.}
{\em For any $D$, $b$ and $a > a_c(D,b)$ equation (\ref{q4}) has a positive stationary solution
vanishing at infinity for all $h_1$ sufficiently small and $h_2$ sufficiently large ($r(h_2)=1$).}

\medskip

\begin{figure}[htbp]
\centerline{\includegraphics[scale=0.3]{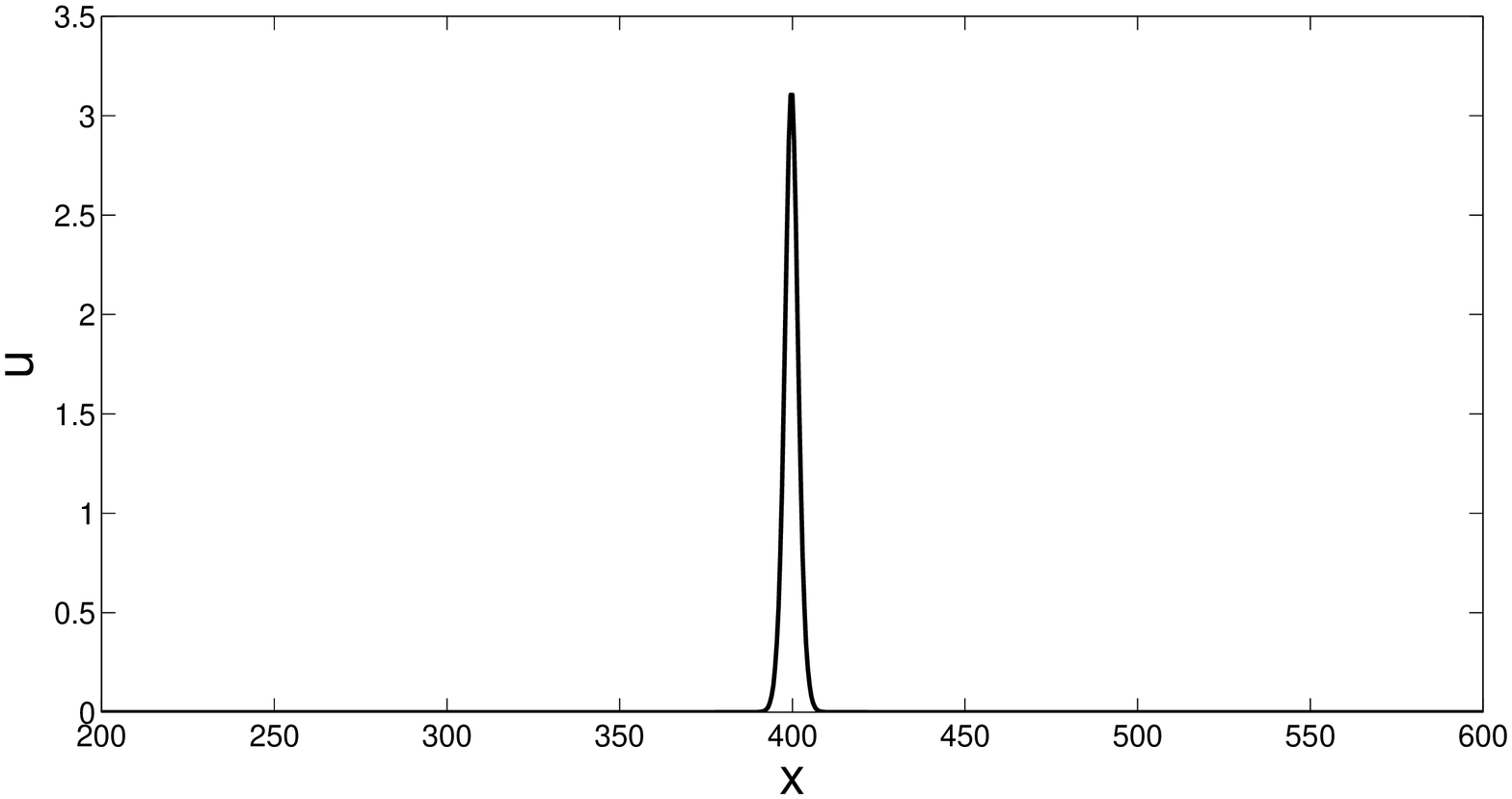}
 \includegraphics[scale=0.3]{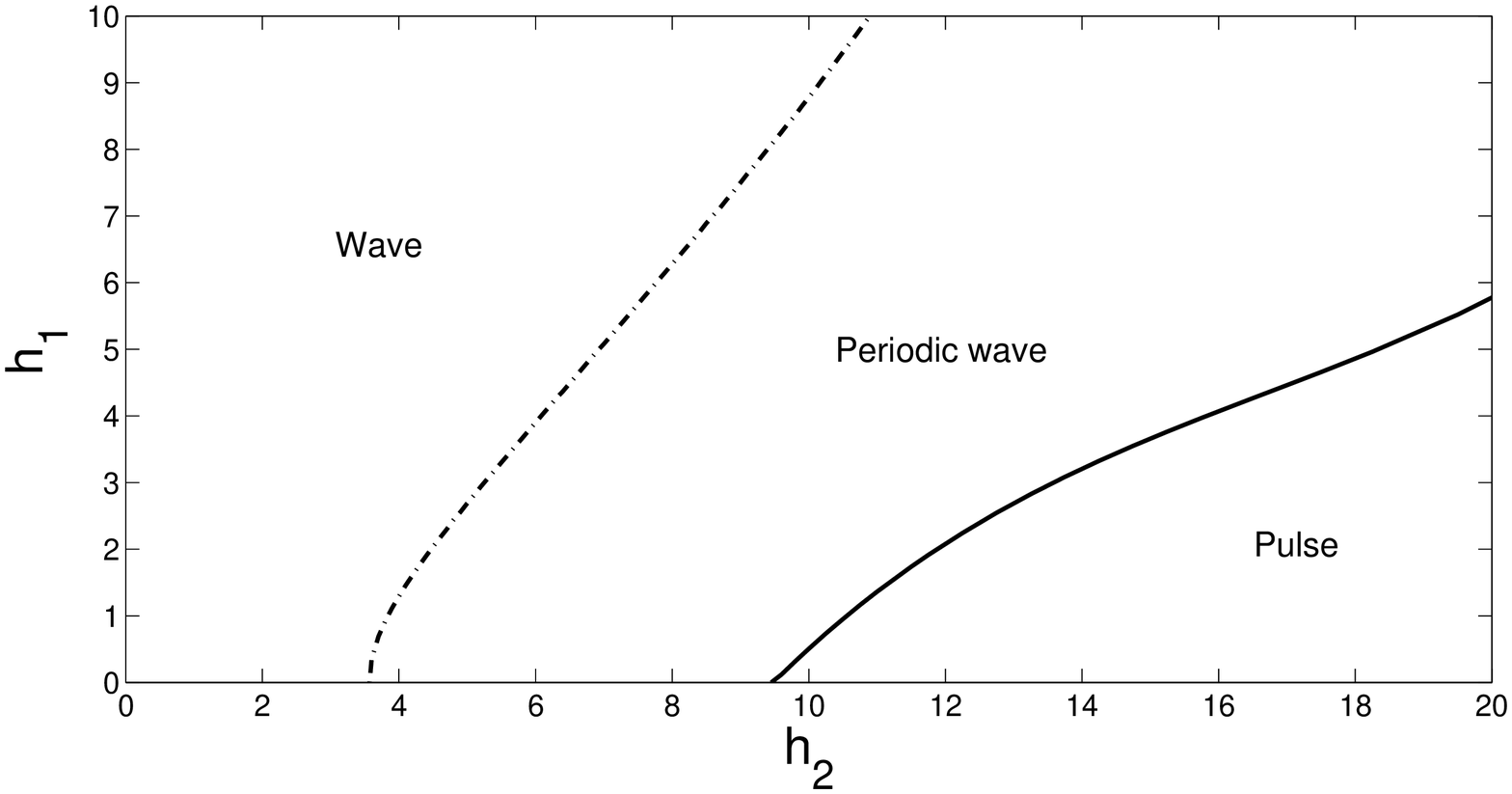}}
 \caption{Pulse solution for equation (\ref{q4}) (left). The regions with different
 regimes on the $(h_2,h_1)$ plane (right).}
 \label{pulse}
\end{figure}

\noindent
The proof of this theorem is similar to the proof of the existence of pulses for equation (\ref{in4}) \cite{VV2}.
It is based on the existence result for equation (\ref{non1}), implicit function theorem and spectral
properties of the linearized operator. Existence of pulses for a system of two equations describing
male and female density distributions is proved in \cite{v3}.
Example of a pulse solution obtained in numerical simulations is shown in Figure \ref{pulse} (left).

\subsection{Transition between different regimes}

Depending on the values of parameters equation (\ref{q4}) can have solutions of different types.
For all other parameters fixed, usual travelling waves (with a constant speed and profile) are
observed for sufficiently small values of $h_2$. Periodic travelling waves exist for intermediate
values and stable pulses for sufficiently large $h_2$. Pulses exist also for other values of $h_2$
but they are unstable. The regions on the $(h_2,h_1)$ plane where simple waves, periodic waves and stable
pulses are observed numerically are shown in Figure \ref{pulse} (right).

Transition from simple to periodic waves occurs due to the essential spectrum crossing the imaginary axis.
This type of bifurcations is discussed in \cite{v2} for equation (\ref{in4}). The homogeneous in space stationary
solution $u=u_-$ loses its stability resulting in appearance of a stationary periodic solution.
The travelling wave connects the constant value $u=u_+$ for $x=+\infty$ with this periodic solution
for $x \to -\infty$. Therefore the wave becomes also periodic.

\begin{figure}[htbp]
\centerline{\includegraphics[scale=0.3]{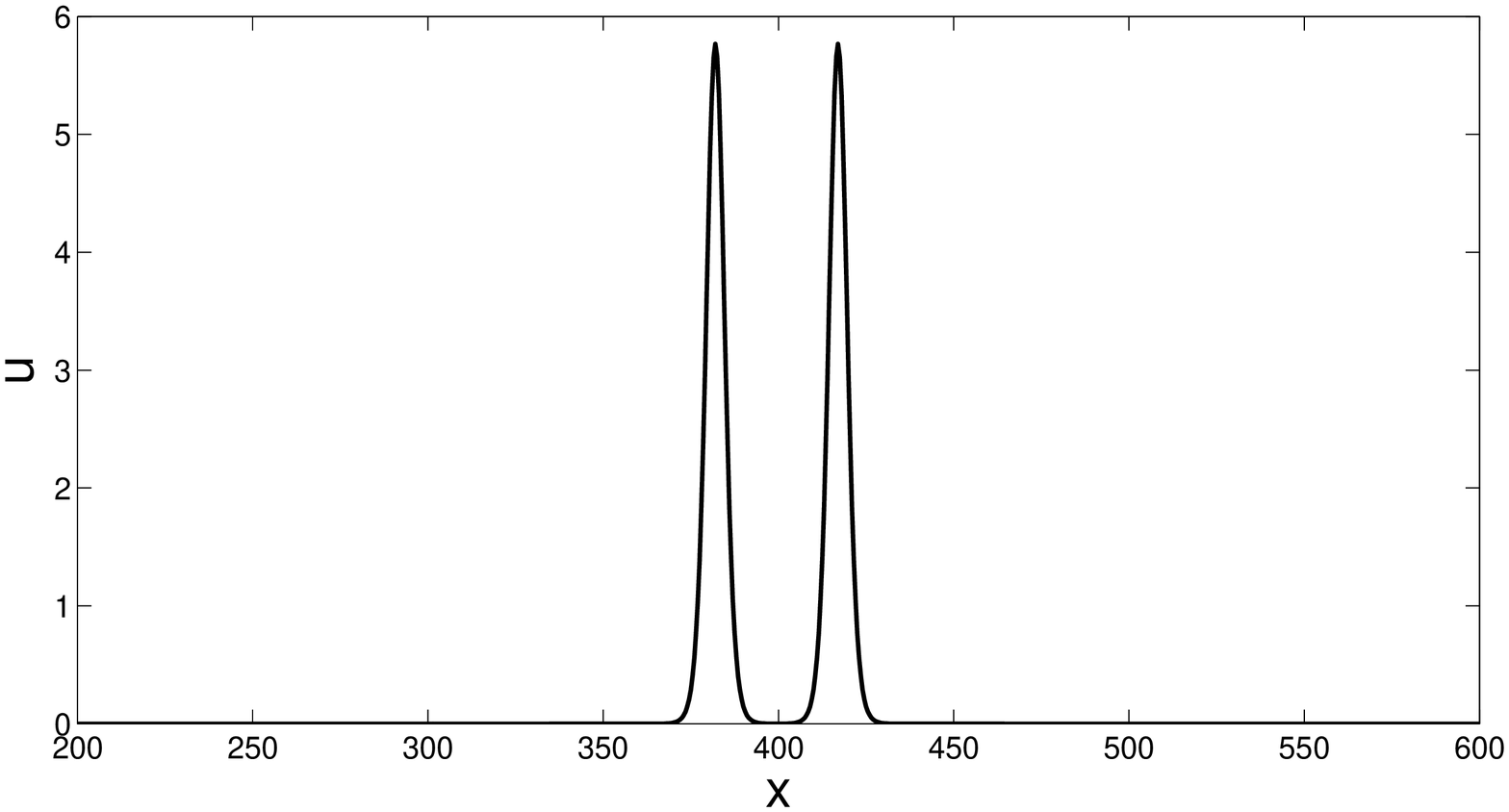}
 \includegraphics[scale=0.3]{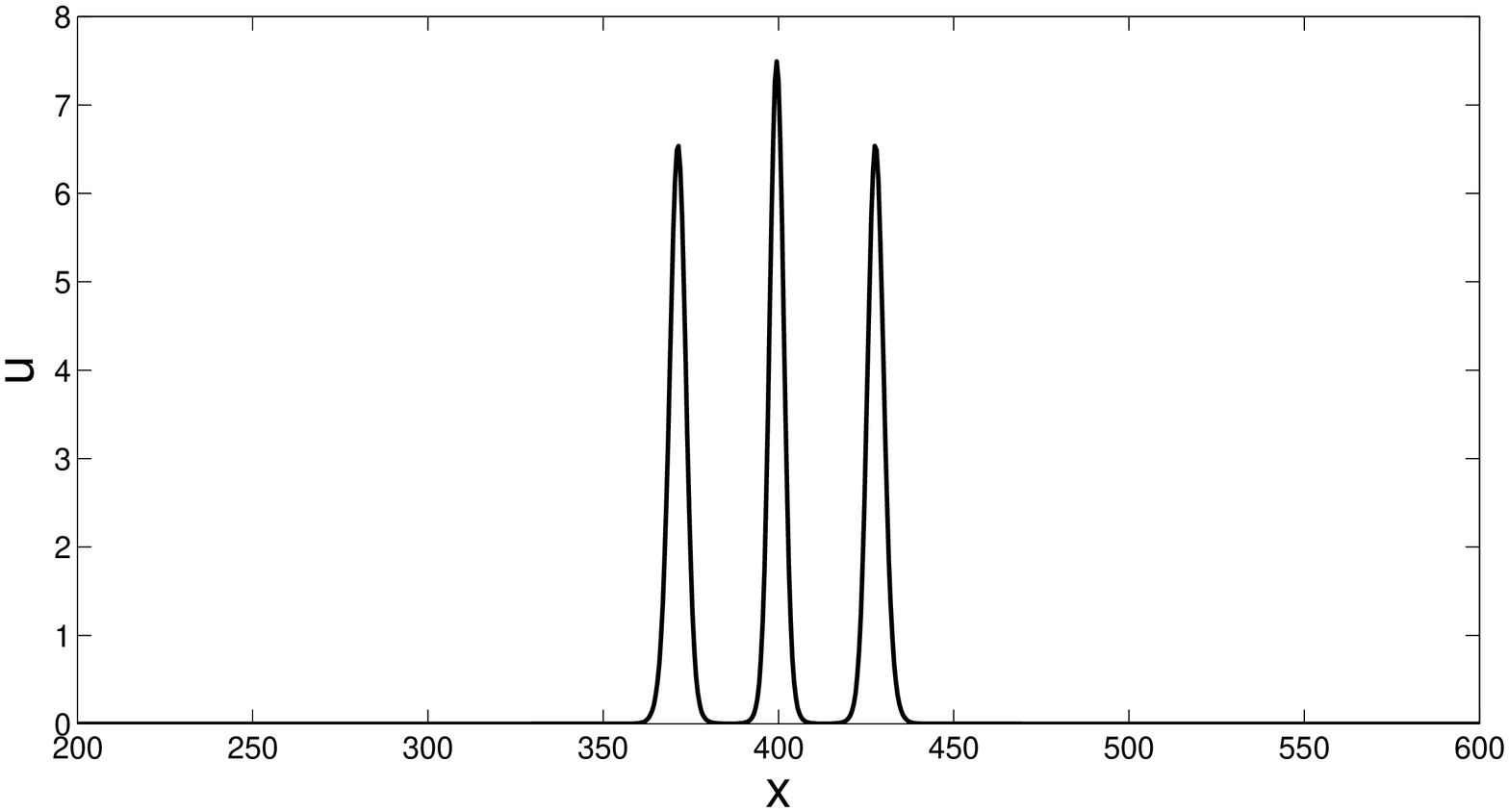}}
 \caption{Solutions of equation (\ref{q4}) with two and three pulses. The pulses slowly
 move from each other.}
 \label{2-3}
\end{figure}

Transition from periodic travelling waves to pulses occurs according to the following scenario.
The average speed of periodic travelling waves decreases when $h_2$ increases. For $h_2$ sufficiently large this
speed becomes zero. Instead of a periodic travelling wave we obtain one or several pulses. If there are more
than one pulse, then they slowly move from each other with a speed decaying in time.
The number of pulses depends on the initial condition (Figure \ref{2-3}).

Both transitions are influenced by the value of $h_1$. Increase of $h_1$ stabilizes simple waves versus
periodic waves, and periodic waves versus pulses.

\section{Discussion}


Nonlocal consumption of resources in reaction-diffusion equations changes dynamics of solutions resulting
in appearance of periodic travelling waves and stable pulses that do not exist for the usual (scalar)
reaction-diffusion equations. In this work we introduce an additional nonlocal term in the reproduction rate.
From the biological point of view, it shows that offsprings can have phenotype different from the phenotype of parents.
If nonlocal consumption of resources destabilizes solutions, the second
nonlocal term stabilizes them. This conclusion concerns the transition from the homogeneous in space solution
to a periodic solution, transition from simple waves to periodic waves, and transition from periodic waves
to stable pulses.

\paragraph{Spectrum, stability, bifurcations.}

The influence of integral terms on stability can be studied analytically for the homogeneous in space stationary solution.
Nonlocal consumption destabilizes this solution leading to the bifurcation of periodic in space solutions due
to a real eigenvalue crossing the origin. 

The situation is different in the case of global consumption where the kernel of the integral identically equals $1$.
The integral $I(u)$ is defined only for integrable functions. Therefore we cannot consider the homogeneous in space
stationary solution on the whole axis if it is different from $0$. In order to study stability of such solutions and the
emergence of pulses, we consider this equation on a bounded interval.

The origins of the instability for nonlocal and global consumption are different.
Consider first the eigenvalue problems (\ref{evproblem}).
Since $\tilde \phi (\xi)=1$ for $\xi=0$ and $\tilde \phi (\xi)$ can be negative for some $\xi \neq 0$,
then the integral term can lead to the instability of solution with some periodicity determined
by this value of $\xi$. In this case the instability occurs due to the principal eigenvalue crossing
the origin from the left half-plane to the right half-plane.

On the other hand, in problem (\ref{5}), $I_0(u)=0$ for $\xi \neq 0$ (admissible $\xi$ are determined by the length $L$ of the interval)
and $I_0(u) \neq 0$ only for $\xi=0$. Hence $I_0(u)$ changes only the principal eigenvalue and does not influence
other eigenvalues. Since $b>0$, then the principal eigenvalue of the operator $Tu = D u'' + bu$ is positive.
The integral $I_0(u)$ moves it to the left half-plane. Hence stability of solutions is determined by the second
eigenvalue of the operator $T$. If it is positive, then the homogeneous in space stationary solution loses
its stability resulting in appearance of the solution with a half period on the interval $L$ ($n=1$).
This solution corresponds to the pulse on the interval $2L$.

Thus, the cases of nonlocal and global consumption differ not only by the origin of instability but also
by periodicity of solutions. For the nonlocal consumption, solution can have several periods on the interval
$[0,L]$ while for the global consumption only half period. If $L$ increases, the period of nonlocal solution can
remain the same while the period of global solution will increase together with $L$. In this limit as $L$
tends to infinity, we obtain a periodic solution in the nonlocal case and a pulse solution in the global case.

Stability and bifurcations of solutions are determined not only by the eigenvalues but also by the essential spectrum.
It can be explicitly determined for the local and nonlocal reaction-diffusion equations \cite{AV1, AV2, v2011, Zhao}.
Transition from simple to periodic waves is related to the essential spectrum crossing the imaginary axis.
Such bifurcations cannot be studied by conventional methods of the bifurcation theory.

Another unusual bifurcation is observed for the transition from periodic waves to pulses. It is a global bifurcation
where peaks of the periodic wave become stationary or slowly moving pulses as the speed of the wave converges to zero.

\paragraph{Emergence of species.}

One of the important applications of reaction-diffusion equations with nonlocal consumption of resources concerns the emergence of species
due to natural selection. In this case, the space variable $x$ corresponds to the phenotype, $t$ is time,
$u(x,t)$ is the density distribution with respect to the phenotype (for each $t$ fixed).

Let us begin this discussion with equation (\ref{in4}). The first term in the right-hand side of this equation
describes small random mutations, the second term reproduction of the population, the last term its mortality.
The reproduction term is proportional to the second power of the population density (sexual reproduction) and
to available resources. Nonlocal consumption of resources $J(u)$ is related to the intraspecific competition.
Indeed, since each individual consumes resources in some area around its average location, then these areas overlap and
the individuals compete for resources. This nonlocal term
leads to the appearance of separated peaks of the density distribution which do not exist in conventional local models.
These peaks in the density distribution correspond to different phenotypes, and we can interpret them as different species.
This interpretation corresponds to Darwin's definition of species as groups of morphologically similar individuals.
From this point of view, appearance of new peaks in the process of propagation of periodic waves
describes the emergence of species \cite{BRV, GVA1, GVA2}. This is sympatric speciation where species are
not separated by genetic or geographic barriers.

Emergence of species in this model is based on three assumptions: random mutations (diffusion), intraspecific competition
(nonlocal consumption) and reproduction with the same phenotype. Let us discuss the last assumption.
The density square in the reproduction term in equation (\ref{in4}) corresponds to the product of densities
of males and females taken at the same space point $x$, that is for the same phenotype.
Therefore this model implies that the phenotypes of parents are the same, and the phenotype
of offsprings is the same as phenotype of parents. This assumption is restrictive and it is not biologically realistic.

Suppose now that phenotypes of males and females in the production term can be different. How this assumption will influence the emergence of species?
Instead of equation (\ref{in4}) we consider now equation (\ref{q4}). In this case, males and females can have different
phenotypes. We suppose that offsprings with phenotype $x$ can be born from males and females with the phenotypes from the interval
$(x-h_1,x+h_1)$. This assumption about the phenotypes can be different. We expect that the qualitative behavior of solutions
will remain the same.

Since the emergence of species in our model is associated with a periodic wave, then we should determine how the integral
$S(u)$ influences the transition from simple to periodic waves. From Figure \ref{pulse} (right) we conclude that periodic waves
exist if $h_1$ is sufficiently small. Hence if the interval of admissible phenotypes of parents is sufficiently large, then the speciation
does not occur.

Furthermore, this figure allows us to conclude that periodic waves appear if $h_1 < h_2$. Let us recall that nonlocal
consumption of resources determines periodicity of solutions. Namely, there is one peak of the population density in the interval
$2 h_2$. Condition $h_1 < h_2$ means that the individuals from the two neighboring peaks (species) cannot have common offsprings;
species are reproductively isolated. The impossibility to have common offsprings can be determined by
morphological or by genetic differences in the individuals and not by some imposed exterior barriers at it is the case in the allopatric  speciation.

This conclusion corresponds to Mayr's definition of species. He suggested that
a species is not just a group of morphologically similar individuals,
but a group of individuals that can breed only among themselves, excluding all others \cite{Mayr}.
Thus we show that species in the sense of Darwin can emerge only if Mayr's condition is satisfied.
In terms of the model considered here this condition can be formulated as $h_1<h_2$.

Emergence of species gives a selective advantage since the total biomass increases \cite{v2014}.
As we discussed above, the species emerge due to the intraspecific competition under the condition of separation of reproduction.
On the other hand, separation of reproduction occurs due to the exclusion of the different from
the process of reproduction.
Therefore we can suppose that exclusion of the different, often observed in nature, is formed in the process of evolution as one
of the mechanisms of speciation.



\bigskip




\bigskip

\appendix


\setcounter{equation}{0}

\section{Appendix}

\subsection{Derivation of the model}

Let $w(x,t)$ and $v(x,t)$ be the density distributions of males and females depending on the phenotype $x$ and time $t$.
 Suppose that males with phenotype $y_1$ and females with phenotype $y_2$ can have offsprings with phenotype $x$ with
 probability $P(x,y_1,y_2)$. Then in the case of unlimited resources the natality rate of offspring with phenotype $x$ is given by the integral

 $$ Q(w,v) = \int_{-\infty}^\infty \int_{-\infty}^\infty P(x,y_1,y_2) w(y_1,t) v(y_2,t) dy_1 dy_2 . $$
Taking into account that natality rate depends on the available resources, we obtain the following equations
for the distributions of males and females:

\begin{equation}\label{q1}
   \frac{\partial w}{\partial t} = D_1 \frac{\partial^2 w}{\partial x^2} + a_1 Q(w,v) (1 - J(w) - J(v)) - b_1 w ,
\end{equation}

\begin{equation}\label{q2}
   \frac{\partial v}{\partial t} = D_2 \frac{\partial^2 v}{\partial x^2} + a_2 Q(w,v) (1 - J(w) - J(v)) - b_2 v ,
\end{equation}
where

$$ J(w) = \int_{-\infty}^\infty \phi(x-y) w(y,t) dy, \;\;\;
J(v) = \int_{-\infty}^\infty \phi(x-y) v(y,t) dy . $$
Assuming that $D_1=D_2$, $a_1=a_2$, $b_1=b_2$, we get $w=v$ (if the initial conditions are equal), and we can reduce this
system to the single equation

\begin{equation}\label{q3}
   \frac{\partial u}{\partial t} = D \frac{\partial^2 u}{\partial x^2} + a Q(u,u) (1 - J(u)) - b u ,
\end{equation}
where $D=D_1$, $a=a_1/2$, $b=b_1$.

In order to study this equation, we need to specify the function $P$.
We will suppose that $P(x,y_1,y_2) = p(x-y_1,x-y_2)$, that is the probability density function
depends on the difference between the phenotypes. We will consider $p(\xi_1,\xi_2)$ as a piece-wise constant function,

\begin{equation}\label{ap1}
  p(\xi_1,\xi_2) = r(N) \times \left\{
  \begin{array}{ccc}
  1 &,& |\xi_1| \leq N \; {\rm and} \; |\xi_2| \leq N \\
  0 &,& |\xi_1| > N \; {\rm or} \; |\xi_2| > N
  \end{array} \right. ,
\end{equation}
where we set $r(N) = 1/N^2$ to get the integral of $p$ equal $1$.
Then $ p(\xi_1,\xi_2) = r(N) \psi(\xi_1) \psi(\xi_2)$, where

$$ \psi(\xi) =  \left\{
  \begin{array}{ccc}
  1 &,& |\xi| \leq N  \\
  0 &,& |\xi| > N
  \end{array} \right. , $$
and $Q(u,u) = (S(u))^2$, where

$$ S(u) = \frac{1}{N} \int_{-\infty}^\infty \psi(x-y) u(y,t) dy . $$
Hence equation (\ref{q3}) can be written as follows:

\begin{equation}\label{ap2}
   \frac{\partial u}{\partial t} = D \frac{\partial^2 u}{\partial x^2} + a (S(u))^2 (1 - J(u)) - b u .
\end{equation}
The particular form of the function $P$ considered above simplifies the analysis of this equation.
We expect that the main properties of solutions remain the same for other functions.

\medskip


\subsection{Positiveness of solutions}

We will begin with a more complete model for the density $u$ of the population and $R$ of resources:

\begin{equation}\label{p1}
  \frac{\partial u}{\partial t} = D \frac{\partial^2 u}{\partial x^2} + a (S(u))^2 R - b u ,
\end{equation}

\begin{equation}\label{p2}
 \epsilon  \frac{dR}{dt} = K - f(R) \int_{-\infty}^\infty \phi(x-y) u(y,t) dy - \sigma R .
\end{equation}
Here the natality term $S^2 R$ is proportional to available resources.
The first term is the right-hand side of equation (\ref{p2}) describes
production of resources with a constant rate. Consumption of resources $f(R) J(u)$ at space
point $x$ depends on available resources through the function $f(R) \geq 0$, $f(0)=0$,
and on the density of individuals $J(u)$ coming from spaces points $y$.

If we formally put $\epsilon=0$, then we get

$$ R = (K-f(R)J(u))/\sigma . $$
We can substitute this expression into (\ref{p1}) but it contains $R$, and this equation does not represent a closed model. We
set

$$ f(R) = \left\{
\begin{array}{ccc}
f_0 &,& R > 0 \\
0   &,& R = 0
\end{array} \right. . $$
This means that the consumption rate by each individual is constant if there are available resources, and it becomes zero
when there are no resources. Then for $R>0$ we get

$$ R = (K - f_0 J(u))/\sigma . $$
Hence equation (\ref{p1}) can now be written as

\begin{equation}\label{p3}
  \frac{\partial u}{\partial t} = D \frac{\partial^2 u}{\partial x^2} + a (S(u))^2 (K - f_0 J(u))/\sigma - b u ,
\end{equation}
assuming that

\begin{equation}\label{p4}
  J(u) < K/f_0 .
\end{equation}
This equation coincides with equation (\ref{q4}) up to the change of variables.

It can be easily verified that if initial condition of system (\ref{p1}), (\ref{p2}) considered on the whole axis is non-negative,
then the solution is also non-negative. Positiveness of solution is preserved for equation (\ref{p3}) if condition (\ref{p4})
is satisfied. If this condition is not satisfied, then positiveness can be violated.
Numerical simulations presented in this work satisfy the positiveness condition.

\end{document}